%% file: main.tex
\documentclass[11pt]{article}
\usepackage[a4paper,margin=1in]{geometry}
\usepackage{amsmath,amssymb}
\usepackage{graphicx}
\usepackage{url}
\usepackage[hidelinks]{hyperref}
\usepackage{booktabs}
\usepackage{microtype}
\usepackage{comment}
\usepackage{placeins}
 \usepackage{color}
\usepackage[acronym,toc,shortcuts]{glossaries} \makeglossaries \input{acro.tex}
\newcommand{\papertitle}{NL-FRA: A MATLAB package for nonlinear frequency response analysis}

\begin{document}
\begin{center}
{\LARGE \papertitle \par}
\vspace{0.45em}
{\large
Rajintha Gunawardena\textsuperscript{1}, Zi-Qiang Lang\textsuperscript{2}, Fei He\textsuperscript{1}*
\par}
\vspace{0.25em}
\textsuperscript{1} Centre for Computational Science and Mathematical Modelling / Coventry University, Coventry CV1 5FB \\
\textsuperscript{2} School of Electrical and Electronic Engineering / The University of Sheffield, Western Bank, Sheffield S10 2TN, UK \\
*correspondence: fei.he@coventry.ac.uk
\end{center}
\vspace{1em}
\noindent \textbf{Software} \\
Repository: \url{https://github.com/raj-gun/NL-FRA} \\
Version: v0.1.0 \\
License: BSD-3-Clause
\vspace{1em}
%============================================
%============================================
\section*{Summary}
Nonlinear Output Frequency Response Functions (NOFRFs) provide a one-dimensional frequency-domain representation of nonlinear dynamics, enabling direct decomposition of an output spectrum into contributions from different orders of nonlinearity. NL-FRA (NonLinear Frequency Response Analysis) is an open-source MATLAB package implementing a data-driven \ac{LS} method for estimating NOFRFs directly from system input–output data. It provides an integrated workflow for NOFRF estimation, validation, and visualisation, together with nonlinear transmissibility analysis. The package supports different types of inputs, i.e. general band-limited, multi-one and harmonic inputs, and includes routines for identifying the frequency ranges over which each nonlinear order contributes to the output. Furthermore, since the \ac{LS} approach estimates NOFRFs directly from input-output data, it can be applied to experimental measurements or data generated using any suitable dynamic model of the system. The software facilitates interpretable nonlinear frequency-domain analysis for applications including fault diagnosis, condition and structural health monitoring, and biomedical engineering.
%
%============================================
%============================================
\section*{Statement of need}
Frequency-domain analysis of linear systems is commonly based on \ac{FRF}s, which uniquely characterise a system’s steady-state dynamics independently of its time-domain representation \cite{ljung1998system}. An \ac{FRF}, i.e. the ratio of output to input spectra, quantifies the response across frequencies and is typically shown as a Bode plot with magnitude and phase versus frequency \cite{dorf2011a}. In the case of \ac{LTI} systems, the output frequency response $Y(j\omega)$, at frequencies $\omega$, is
\begin{equation}\label{eq:FRF}
     Y(j\omega) = G(j\omega)U(j\omega), 
\end{equation}
where $Y(j\omega)$ and $U(j\omega)$ are the input and output spectra, respectively. $G(j\omega)$ is the \ac{FRF}. In an LTI system, the output contains only frequency components present in the input \cite{ljung1998system, dorf2011a}. By contrast, nonlinear systems can generate additional frequency components, including harmonics, sub-harmonics, and intermodulation products, that are absent from the input through energy transfer between frequencies \cite{lang2005energy, peng2007b, powers1994a, lang1997a, lang1996a, he2021nonlinear}.

Lang and Billings \cite{lang1996a} derived that the output spectrum or output frequency response, $Y(\omega)$, of a nonlinear system can be decomposed as
\begin{equation}\label{eq:OFR_NL}
    \begin{cases}
            \ \ \ \ \ \ \ \ \ \ \ \ \ \ \ \ \ \ \ \ \ \ \ \
            \displaystyle Y(\omega) = \sum_{n=1}^{N} Y_n(\omega)\ \ \ \forall\omega\\[15pt]
            %%%%%%%%%
            \displaystyle
            Y_n(\omega) = \frac{ 1 / \sqrt{n}  }{ \left( 2\pi \right)^{n-1} } \int_{\omega} H_n \left( \omega_1 , \cdots , \omega_n \right) \prod_{i=1}^{n} U( \omega_i) d\sigma_{n\omega},
    \end{cases}
\end{equation}
where $U(\omega)$ denotes the spectrum of the input signal at the frequency $\omega$. $H_n \left( \omega_1 , \cdots , \omega_n \right)$ is known as the $n$\textsuperscript{th} order \ac{GFRF} \cite{George1959}. $Y_n(\omega)$ is the output frequency response of the $n$\textsuperscript{th} order. For $n=1$, $H_1(\omega_1)$ is a \ac{FRF} that describes the linear dynamics; $Y_1(\omega)$ denotes the output spectrum produced by the linear dynamics of the nonlinear system. Similarly, $H_n \left( \ \right)$, for $n\geq2$, describes the higher-order spectral characteristics of the $n$\textsuperscript{th} order nonlinearity, while $Y_n(\omega)$ denotes the output spectrum produced by the $n$\textsuperscript{th} order nonlinearity. %It is clear from comparing Eqs. \eqref{eq:FRF} and \eqref{eq:OFR_NL} how intricate nonlinear dynamics are.

However, the \ac{GFRF}s are multi-dimensional, computationally heavy, and difficult to interpret. Lang and Billings \cite{lang2005energy} introduced the concept of \ac{NOFRF}s to elucidate the energy transfer phenomenon in nonlinear systems and provided a more accessible representation of $Y_n(j\omega)$ in Eq. \eqref{eq:OFR_NL}, transforming multi-dimensional frequency-response information to one-dimensional functions of frequency as
\begin{equation}\label{eq:OFR_NOFRF}
    \begin{cases}
        \displaystyle Y(\omega) = \sum_{n=1}^{N} Y_n(\omega)\ \ \ \forall\omega\\[15pt]
        
        \displaystyle Y_n(j\omega) = G_n(j\omega)U_n(j\omega).
    \end{cases}
\end{equation}
$Y_n(j\omega)$ is the $n$\textsuperscript{th}-order \ac{OFRF}. $G_n(j\omega)$ is the $n$\textsuperscript{th}-order \ac{NOFRF},
\begin{equation}\label{eq:NOFRF}
    G_n(\omega) = \frac{ \int_{ \omega } H_n \left( \omega_1 , \cdots , \omega_n \right) \prod_{n}^{i=1} U( \omega_i) \ \ d\sigma_{n\omega} }{  \int_{ \omega }\prod_{n}^{i=1} U( \omega_i) \ \ d\sigma_{n\omega}  }.
\end{equation}
Here, the \ac{NOFRF}s are normalised \ac{GFRF}s that are input-dependent. Furthermore, $G_n(\omega)$ only exists, when $\int_{\omega}\prod_{n}^{i=1} U( \omega_i) \ \ d\sigma_{n\omega} \neq 0$. In Eq. \eqref{eq:OFR_NOFRF}, $U_n(j\omega)$ is the $n$\textsuperscript{th}-order composition of the input $U(j\omega)$ and is given as,
\begin{equation}\label{eq:U_n}
    U_n(\omega) = \frac{ 1 / \sqrt{n}  }{ \left( 2\pi \right)^{n-1} } \int_{ \omega = \omega_1 + \cdots + \omega_n } \prod_{n}^{i=1} U( \omega_i) \ \ d\sigma_{n\omega}.
\end{equation}
Here, $\int_{\omega}\prod_{n}^{i=1} U( \omega_i) \ \ d\sigma_{n\omega} = FT\{u^{n}(t)\}$, where $FT\{ \cdot \}$ denotes the Fourier transform. Therefore,
\begin{equation} \label{eq:U_n_FFT}
U_n(j\omega) = \frac{1/\sqrt{n}}{(2\pi)^{n-1}} FT\{u^{n}(t)\},
\end{equation}
From the above, it can be simply stated that at frequencies $\omega$, where $U_n(j\omega) = 0$, $G_n(j\omega)$ does not exist and $Y_n(j\omega)$ doesn't contribute to $Y(j\omega)$. Two evaluation methods are available to evaluate \ac{NOFRF}s: (1) an associated linear equations (ALE) method (this is a symbolic method) that relies on differential- or difference-equation models \cite{bayma2018}, and (2) a data-driven least-squares \ac{LS} based method that estimates \ac{NOFRF}s from input–output data \cite{Gunawardena2018, lang2005energy}. \ac{NOFRF}s have been used for structural damage detection and health monitoring \cite{Peng2007Crack}, fault diagnosis and condition monitoring of rotor-bearing systems \cite{Zhu2018Rotor}, tool-condition monitoring in manufacturing \cite{Zhu2022NOFRF}, and fault diagnosis of electromechanical drives \cite{Chen2020Fault}. To the best of our knowledge, however, no open-source package implements either method. This package focuses on the \ac{LS} approach, since any valid dynamic representation of a system (even a neural network \cite{Jacobs2024}) can provide the input–output data needed to evaluate \ac{NOFRF}s. Thus, using the data-driven \ac{LS} method, along with an appropriate probing input, one can visualise and analyse the multi-dimensional higher-order spectral characteristics of any nonlinear system using one-dimensional frequency functions. 

Since \ac{NOFRF}s are invariant to input scaling, the input $u(t)$ can be multiplied by several factors $\alpha_m$ to produce $\alpha_m u(t)$, yielding corresponding outputs $y^{(m)}(t)$ for $m=1,\ldots,M$. Evaluate the spectra $Y^{(m)}(j\omega)=FT\{ y^{(m)}(t) \}$ for $m=1,\ldots,M$ and $U_n(j\omega)$ from Eq. \eqref{eq:U_n_FFT}. These can then be used to estimate the \acp{NOFRF} via the \ac{LS} approach by incorporating them into a matrix representation of Eq. \eqref{eq:OFR_NOFRF}. This is given by
\begin{equation}
    \mathbf{Y}(j\omega)=\mathbf{AU}(j\omega) \mathbf{G}(j\omega)
\end{equation}
where 
\begin{equation}
    \mathbf{AU}(j\omega) = \left[\mathbf{au}_1(j\omega), \cdots, \mathbf{au}_N(j\omega) \right], 
\end{equation}
\begin{equation}
    \mathbf{G}(j\omega) = \left[ G_1(j\omega), \cdots, G_N(j\omega) \right]^T,
\end{equation}
\begin{equation}
    \mathbf{Y}(j\omega) = \left[ Y^{(1)}(j\omega), \cdots, Y^{(M)}(j\omega) \right]^T
\end{equation}
and
\begin{equation}
    \mathbf{au}_n(j\omega)=
    \begin{bmatrix}
        \alpha_1^{n} U_n(j\omega)\\
        \vdots\\
        \alpha_M^{n} U_n(j\omega)
    \end{bmatrix}.
\end{equation}
However, as mentioned earlier $\forall\omega: \left(U_n(j\omega) = 0 \right) \Rightarrow  \nexists G_n(j\omega)$ resulting in a sparse ill-conditioned information matrix $\mathbf{AU}(j\omega)$. As emphasised in \cite{Gunawardena2018}, this can be avoided by forming an appropriate dense information matrix $\overline{\mathbf{AU}}(j\omega)$ containing only the relevant $\mathbf{au}_n(j\omega)$ terms. For this the frequencies $\Omega_n = \{ \omega|U_n(j\omega) \neq 0 \}$ for $n=1,\cdots,N$ needs to be found. Therefore, at the frequencies $\Omega_n$, $\overline{\mathbf{AU}}(j\omega)$ will only contain the corresponding $\mathbf{au}_n(j\omega)$ terms where $U_n(j\omega) \neq 0$. Once $\overline{\mathbf{AU}}(j\omega)$ is formed, the pseudo-inverse can be used to find the relevant \ac{NOFRF}s, where 
\begin{equation}
    \widehat{\mathbf{G}}(j\omega) = \big[ \overline{\mathbf{AU}}(j\omega)^{\!T} \  \overline{\mathbf{AU}}(j\omega) \big]^{-1} \ \overline{\mathbf{AU}}(j\omega)^{\!T} \ \mathbf{Y}(j\omega).
\end{equation}
For example, lets assume that at $\omega'$ only the 3\textsuperscript{rd} and 4\textsuperscript{th} orders nonlinearities contribute to the final output. In this case, $\mathbf{Y}(j\omega')=\overline{\mathbf{AU}}(j\omega') \mathbf{G}(j\omega')$ where $\overline{\mathbf{AU}}(j\omega') = \left[\mathbf{au}_3(j\omega'), \mathbf{au}_4(j\omega') \right]$ and $\mathbf{G}(j\omega') = \left[ G_3(j\omega), G_4(j\omega) \right]^T$.
% 
%============================================
%============================================
\section*{Features in NL-FRA}
NL-FRA can be used to construct a complete \ac{NOFRF} evaluation and validation pipeline. It also features a pipeline for nonlinear transmissibility analysis across different orders of nonlinearity \cite{Peng2010Transmissibility}, together with relevant plotting functions for \ac{NOFRF}-based transmissibility analysis.

The \ac{NOFRF}s are input-dependent frequency functions. Once a dynamic model of the actual system is obtained, the model can be simulated using an appropriate probing input at different amplitudes to characterise higher-order nonlinear dynamics. Depending on the probing input, the frequency sets $\Omega_n = \{ \omega|U_n(j\omega) \neq 0 \}$ for $n=1,\cdots,N$ needs to be determined. In general, three types of standard probing inputs can be used: (1) sinusoidal, (2) general band-limited input, and (3) multi-tone input. For each of these probing signals, algorithms are available to determine $\Omega_n$ for $n=1,\cdots,N$, which have been implemented within NL-FRA. 

The example below presents results for three cases: (1) \ac{NOFRF}s under band-limited input for an identified \ac{NARX} model using NonSysID \cite{Gunawardena2025}, (2) \ac{NOFRF}s under multi-tone input, and (3) transmissibility analysis using \ac{NOFRF}s, with sinusoidal probing. The GitHub repository provides scripts for reproducing these examples, together with documentation on the use of NL-FRA.
%============================================
%============================================
\section*{Examples}
%============================================
\subsection*{General band-limited input}

\begin{figure}[htbp]
    \centering
    \includegraphics[width=0.8\textwidth]{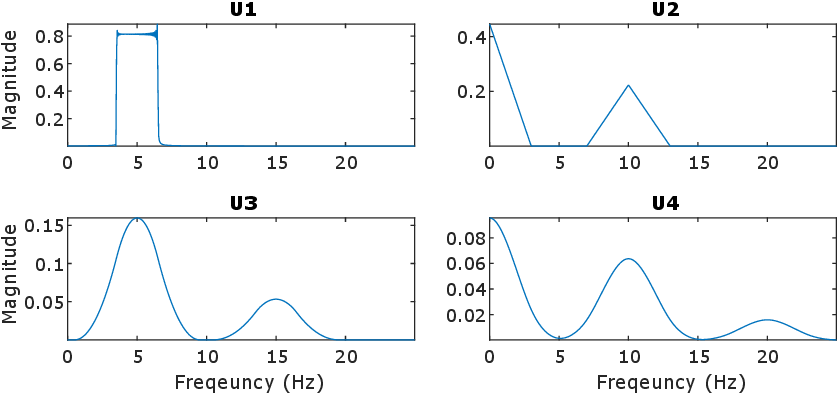}
    \caption{The nonlinear input compositions, $U_n(j\omega)$, up to $n=4$, for the general band-limited input.}
    \label{fig:lang_un}
\end{figure}

\begin{figure}[htbp]
    \centering
    \includegraphics[width=0.8\textwidth]{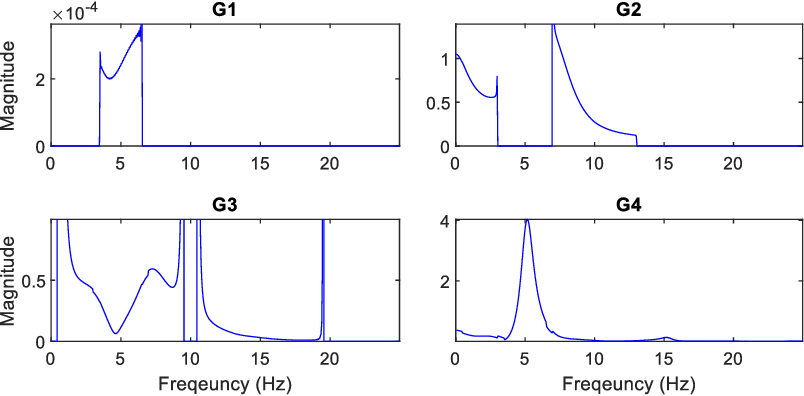}
    \caption{The first four \ac{NOFRF}s, $G_n(j\omega)$, evaluated from the identified NARX model using the general band-limited input.}
    \label{fig:lang_nofrf}
\end{figure}

\begin{figure}[htbp]
    \centering
    \includegraphics[width=0.8\textwidth]{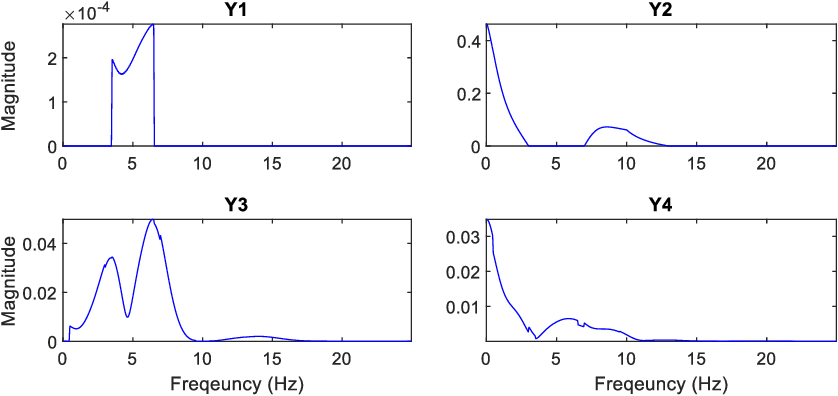}
    \caption{The first four \ac{OFRF}s, $Y_n(j\omega)$, evaluated from the identified NARX model using the general band-limited input.}
    \label{fig:lang_ofrf}
\end{figure}

The first example demonstrates a data-driven workflow using a coupled electric drive system \cite{Wigren2017CED}. A polynomial NARX model is identified (using NonSysID \cite{Gunawardena2025}) from the first uniformly distributed input--output dataset and independently validated using the second dataset. The identified model, with a sampling period of $T_s=0.02$s, is
\begin{equation}\label{eq:lang_example}
\begin{aligned}
y(k)={}&1.7569y(k-1)-1.1078y(k-2)+0.2563y(k-4)-0.049809y(k-7)+0.042299y(k-9)\\
&+0.022931y(k-1)u(k-2)-0.021965y(k-1)u(k-4)+0.013449y(k-3)u(k-3)\\
&-0.015884y(k-7)u(k-2)+0.016512y(k-9)u(k-6)+0.0088507u(k-3)u(k-4)\\
&-0.0066702u(k-3)u(k-5)+0.022401u(k-4)u(k-7)+0.013560u^2(k-5).
\end{aligned}
\end{equation}
The identified NARX model is then simulated using the general band-limited probing input
\begin{equation}\label{eq:lang_input}
u(t)=\frac{3}{2\pi}\frac{\sin(2\pi 6.5t)-\sin(2\pi 3.5t)}{t},
\end{equation}
with its amplitude normalised relative to the identification input. The probing input is scaled at several amplitudes and the resulting simulated input--output data are used to evaluate the first four \ac{NOFRF}s.
\begin{figure}[htbp]
    \centering
    \includegraphics[width=0.8\textwidth]{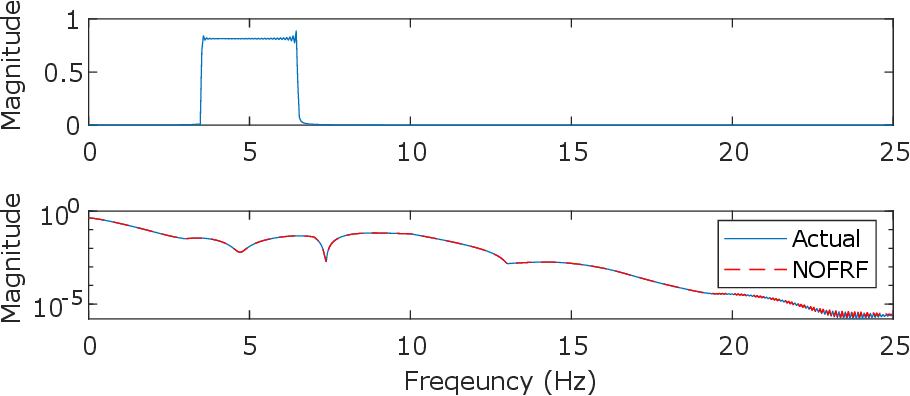}
    \caption{Input and output spectra generated by the identified NARX model using the general band-limited probing input in Eq. \eqref{eq:lang_input}.}
    \label{fig:lang_io}
\end{figure}
\begin{figure}[htbp]
    \centering
    \includegraphics[width=0.8\textwidth]{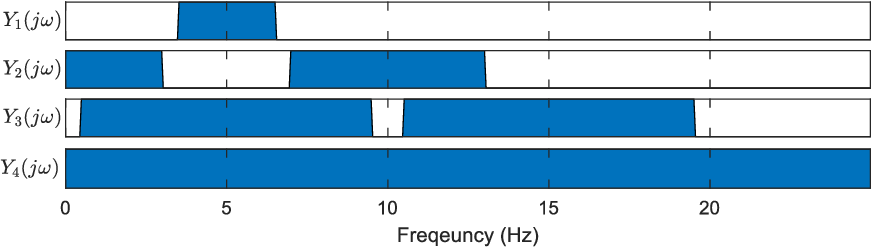}
    \caption{Frequency spaces in which $Y_n(j\omega)$ contributes to the final output $Y(j\omega)$, corresponding to the frequencies where $U_n(j\omega)$ is non-zero for the general band-limited input in Eq. \eqref{eq:lang_input}.}
    \label{fig:lang_nofrf_valid}
\end{figure}
The \texttt{SISO\_NOFRF} function, for the given input type, determines the valid frequency regions for each nonlinear order and evaluates the corresponding \ac{NOFRF}s from the simulated input--output data. Figures \ref{fig:lang_io}--\ref{fig:lang_ofrf} show the input--output response, the input compositions $U_n(j\omega)$, the evaluated \ac{NOFRF}s $G_n(j\omega)$ and \ac{OFRF}s $Y_n(j\omega)$.
\FloatBarrier
%============================================
\subsection*{Multi-tone input}

\begin{figure}[htbp]
    \centering
    \includegraphics[width=0.75\textwidth]{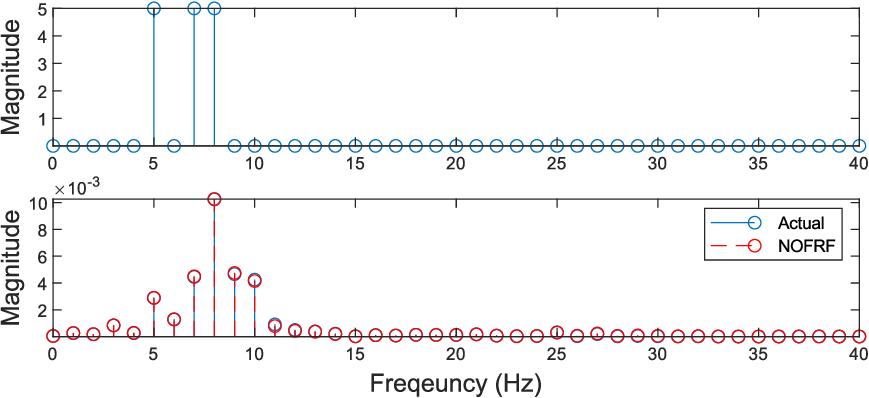}
    \caption{Input and output spectra for the three-tone probing input in Eq. \eqref{eq:multitone_input}.}
    \label{fig:multitone_io}
\end{figure}

\begin{figure}[htbp]
    \centering
    \includegraphics[width=\textwidth]{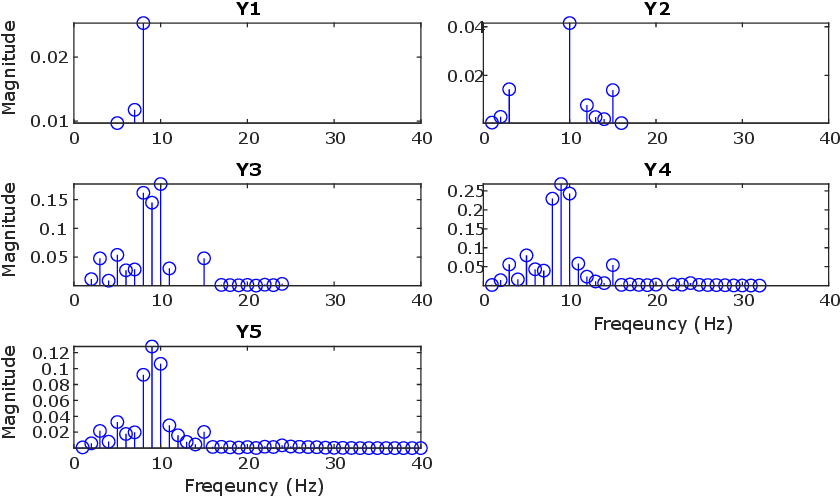}
    \caption{The \ac{OFRF} contributions, $Y_n(j\omega)=G_n(j\omega)U_n(j\omega)$, for the three-tone Duffing example.}
    \label{fig:multitone_ofrf}
\end{figure}

A discrete multi-tone probing input is demonstrated using a Duffing oscillator,
\begin{equation}\label{eq:multitone_system}
\ddot y(t)+0.96\pi\dot y(t)+(12\pi)^2y(t)+0.1(12\pi)^6y^3(t)=u(t),
\end{equation}
where
\begin{equation}\label{eq:multitone_input}
u(t)=A\left[\cos(2\pi5t)+\cos(2\pi7t)+\cos(2\pi8t)\right].
\end{equation}
The \ac{NOFRF}s are evaluated using constant input scalings between 1.3 and 1.2 and are tested at $A=1.4$. The \texttt{SISO\_NOFRF\_comp} function determines the discrete frequency components associated with each nonlinear order and evaluates the corresponding \ac{NOFRF}s. The input and output spectra, evaluated \ac{NOFRF}s, and corresponding \ac{OFRF}s are shown in Figures \ref{fig:multitone_io}--\ref{fig:multitone_ofrf}.

\FloatBarrier
%============================================
\subsection*{Transmissibility analysis}

The transmissibility example uses the Duffing oscillator
\begin{equation}\label{eq:trans_system}
    \ddot y(t)+5.28\pi\dot y(t)+(12\pi)^2y(t)+0.1(12\pi)^6y^3(t)=A\cos(2\pi f t).
\end{equation}
For $\omega=2\pi f$, the fundamental and third-harmonic transmissibilities are \cite{Peng2010Transmissibility, Ho2012Transmissibility}
\begin{equation}\label{eq:trans_def}
    |\mathrm{Trans}(\omega)|=\frac{|Y(j\omega)|}{|U(j\omega)|}, \qquad
    |\mathrm{Trans}_3(3\omega)|=\frac{|Y(j3\omega)|}{|U(j\omega)|}.
\end{equation}
For each excitation frequency, local \ac{NOFRF}s are evaluated from input scalings between 1.3 and 1.2 and used to reconstruct the response at $A=1.4$. 
\begin{figure}[htbp]
    \centering
    \includegraphics[width=\textwidth]{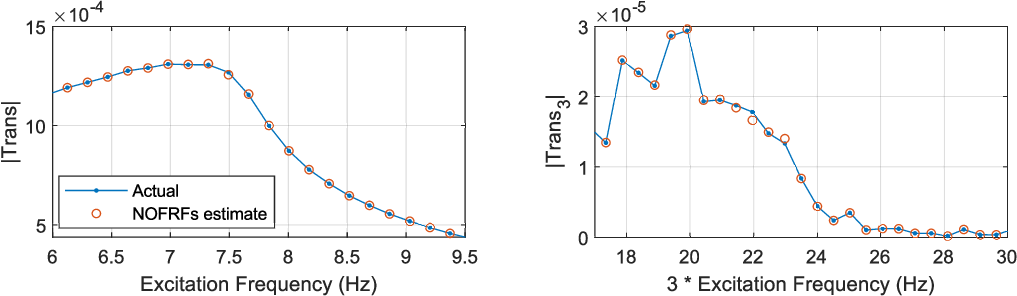}
    \caption{Comparison of the actual and \ac{NOFRF}-generated transmissibilities at the excitation frequency and the third harmonic.}
    \label{fig:trans_comp}
\end{figure}
\begin{figure}[htbp]
    \centering
    \includegraphics[width=\textwidth]{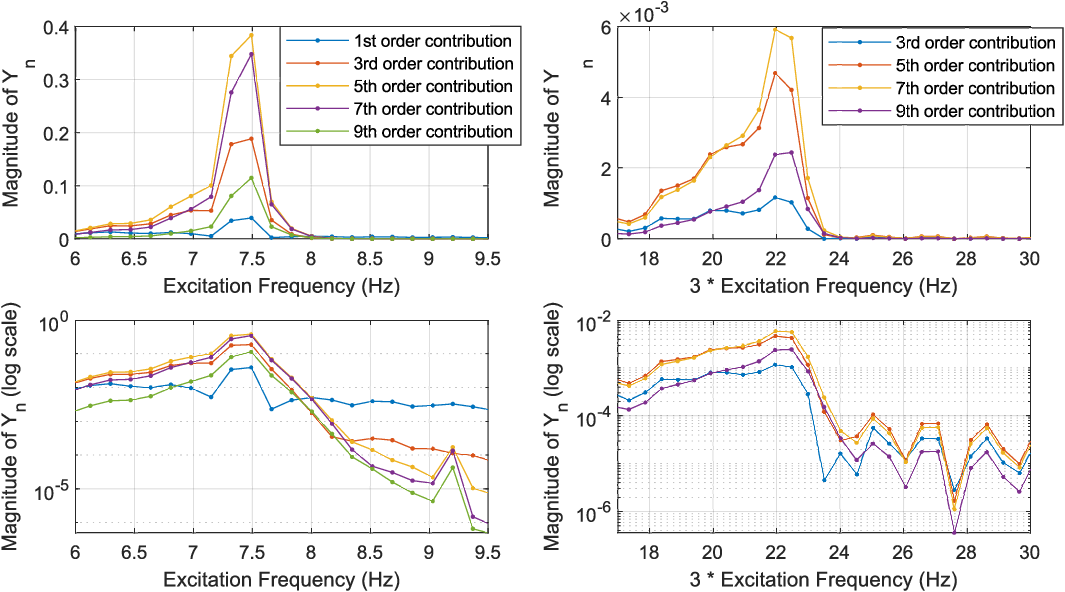}
    \caption{Contributions of the individual nonlinear orders to the output response at the excitation frequency and the third harmonic over the frequency sweep.}
    \label{fig:trans_orders}
\end{figure}
The \texttt{SISO\_NOFRF} function is used to evaluate the local \ac{NOFRF}s at each excitation frequency, while \texttt{nofrf\_test} reconstructs the output spectrum and individual nonlinear-order contributions at the test amplitude. The actual and \ac{NOFRF}-generated transmissibilities are compared at the excitation frequency and its third harmonic in Figure \ref{fig:trans_comp}. The individual nonlinear-order contributions are shown in Figure \ref{fig:trans_orders}.
\FloatBarrier
%============================================
%============================================
\section*{Future work}

NL-FRA currently supports the analysis of \ac{SISO} nonlinear systems. Future updates will extend the package to \ac{MISO} and subsequently \ac{MIMO} systems. Planned developments also include procedures for evaluating \ac{GFRF}s and an open-source Python implementation.

%============================================
%============================================
\section*{Acknowledgements}
RG and FH were supported by EPSRC grant [EP/X020193/1]
%============================================
%============================================

%\newpage
\bibliographystyle{ieeetr}
\bibliography{citations.bib}

\end{document}

%% file: acro.tex
\newacronym{NARX}{NARX}{Nonlinear Auto-Regressive with eXogenous input}
\newacronym{LS}{LS}{Least Squares} 
\newacronym{GFRF}{GFRF}{Generalised Frequency Response Function} 
\newacronym{NOFRF}{NOFRF}{Nonlinear Output Frequency Response Function}
\newacronym{OFRF}{OFRF}{Output Frequency Response Function} 
\newacronym{LTI}{LTI}{Linear time-invariant} 
\newacronym{FRF}{FRF}{Frequency Response Function}
\newacronym{ARX}{ARX}{Auto-Regressive with eXogenous input}
\newacronym{SISO}{SISO}{single-input single-output}
\newacronym{MISO}{MISO}{multi-input single-output}
\newacronym{MIMO}{MIMO}{multi-input multi-output}
\newacronym{EC}{EC}{Eyes-close}
\newacronym{EO}{EO}{Eyes-open}
\newacronym{MRI}{MRI}{Magnetic Resonance Imaging}
\newacronym{fMRI}{fMRI}{functional MRI}
\newacronym{rsfMRI}{rsfMRI}{resting-state fMRI}
\newacronym{EEG}{EEG}{Electroencephalogram}
\newacronym{MEG}{MEG}{Magnetoencephalography}
\newacronym{AD}{AD}{Alzheimer's disease}
\newacronym{PD}{PD}{Parkinson's disease}
\newacronym{HC}{HC}{Healthy Control}
\newacronym{FC}{FC}{Functional Connectivity}
\newacronym{CFC}{CFC}{Cross-Frequency Coupling}
\newacronym{CF}{CF}{Cross-Frequency}
\newacronym{KDE}{KDE}{Kernel Density Estimator}
\newacronym{KNN}{KNN}{\textit{k}-nearest-neighbour estimators}
\newacronym{PDF}{PDF}{Probability Density Function}
\newacronym{SC}{SC}{Structural Connectivity}
\newacronym{EFC}{EFC}{Effective Connectivity}
\newacronym{GC}{GC}{Granger Causality}
\newacronym{MCI}{MCI}{Mild Cognitive Impairment}
\newacronym{CT}{CT}{Computerised Tomography}
\newacronym{PET}{PET}{Positron Emission Tomography}
\newacronym{SVM-MCV}{SVM-MCV}{Linear SVM classification with Monte-Carlo cross-validation}